\documentclass[12pt]{article}
\usepackage{graphicx}

\begin{document}

\title{Complete Experiments for Pion Photoproduction}


\author{Lothar Tiator\\
Institut f\"ur Kernphysik, Johannes Gutenberg-Universit\"at\\
D-55099 Mainz, Germany}

\maketitle

\begin{abstract}
The possibilities of a model-independent partial wave analysis for
pion, eta or kaon photoproduction are discussed in the context of
`complete experiments'. It is shown that the helicity amplitudes
obtained from at least 8 polarization observables including beam,
target and recoil polarization can not be used to analyze nucleon
resonances. However, a truncated partial wave analysis, which
requires only 5 observables will be possible with minimal model
assumptions.
\end{abstract}

\section{Introduction}

Around the year 1970 people started to think about how to determine
the four complex helicity amplitudes for pseudoscalar meson
photoproduction from a complete set of experiments. In 1975 Barker,
Donnachie and Storrow~\cite{Barker75} published their classical
paper on `Complete Experiments'. After reconsiderations and careful
studies of discrete ambiguities~\cite{FTS92,Keaton:1996pe,chiang},
in the 90s it became clear that such a model-independent amplitude
analysis would require at least 8 polarization observables which
have to be carefully chosen. There are plenty of possible
combinations, but all of them would require a polarized beam and
target and in addition also recoil polarization measurements.
Technically this was not possible until very recently, when
transverse polarized targets came into operation at Mainz, Bonn and
JLab and furthermore recoil polarization measurements by nucleon
rescattering has been shown to be doable. This was the start of new
efforts in different groups in order to achieve the complete
experimental information and a model-independent partial wave
analysis~\cite{Workman:2010xc,Workman:2011hi,Dey:2010fb,Sandorfi:2010uv}.

\section{Complete experiments}

A complete experiment is a set of measurements which is sufficient to predict
all other possible experiments, provided that the measurements are free of
uncertainties. Therefore it is first of all an academic problem, which can be
solved by mathematical algorithms. In practise, however, it will not work in
the same way and either a very high statistical precision would be required,
which is very unlikely, or further measurements of other polarization
observables are necessary. Both problems, first the mathematical problem but
also the problem for a physical experiment can be studied with the help of
state-of-the-art models like MAID or partial wave analyses (PWA) like SAID.
With high precision calculations the complete sets of observables can be
checked and with pseudo-data, generated from models and PWA, real experiments
can be simulated under realistic conditions.

\subsection{Coordinate Frames}

Experiments with three types of polarization can be performed in
meson photoproduction: photon beam polarization, polarization of the
target nucleon and polarization of the recoil nucleon. Target
polarization will be described in the frame $\{ x, y, z \}$, see
Fig.~1, with the $z$-axis pointing into the direction of the photon
momentum $\hat{ \vec k}$, the $y$-axis perpendicular to the reaction
plane, ${\hat{ \vec y}} = {\hat{ \vec k}} \times {\hat{ \vec q}} /
\sin \theta$, and the $x$-axis is given by ${\hat{\vec x}} = {\hat{
\vec y}} \times {\hat{\vec z}}$. For recoil polarization,
traditionally the frame $\{ x', y', z' \}$ is used, with the
$z'$-axis defined by the momentum vector of the outgoing meson
${\hat{\vec q}}$, the $y'$-axis is the same as for target
polarization and the $x'$-axis given by ${\hat{\vec{x}'}} =
{\hat{\vec{y}'}} \times {\hat{\vec{z}'}}$.

The photon polarization can be linear or circular. For a linear
photon polarization $(P_T=1)$ in the reaction plane $(\hat{\vec
x},\hat{\vec z})$, $\varphi=0$. Perpendicular, in direction ${\hat{
\vec y}}$, the polarization angle is $\varphi=\pi/2$. Finally, for
right-handed circular polarization, $P_{\odot}=+1$.

\begin{figure}[htb]
\centerline{
\includegraphics[width=0.80\textwidth]{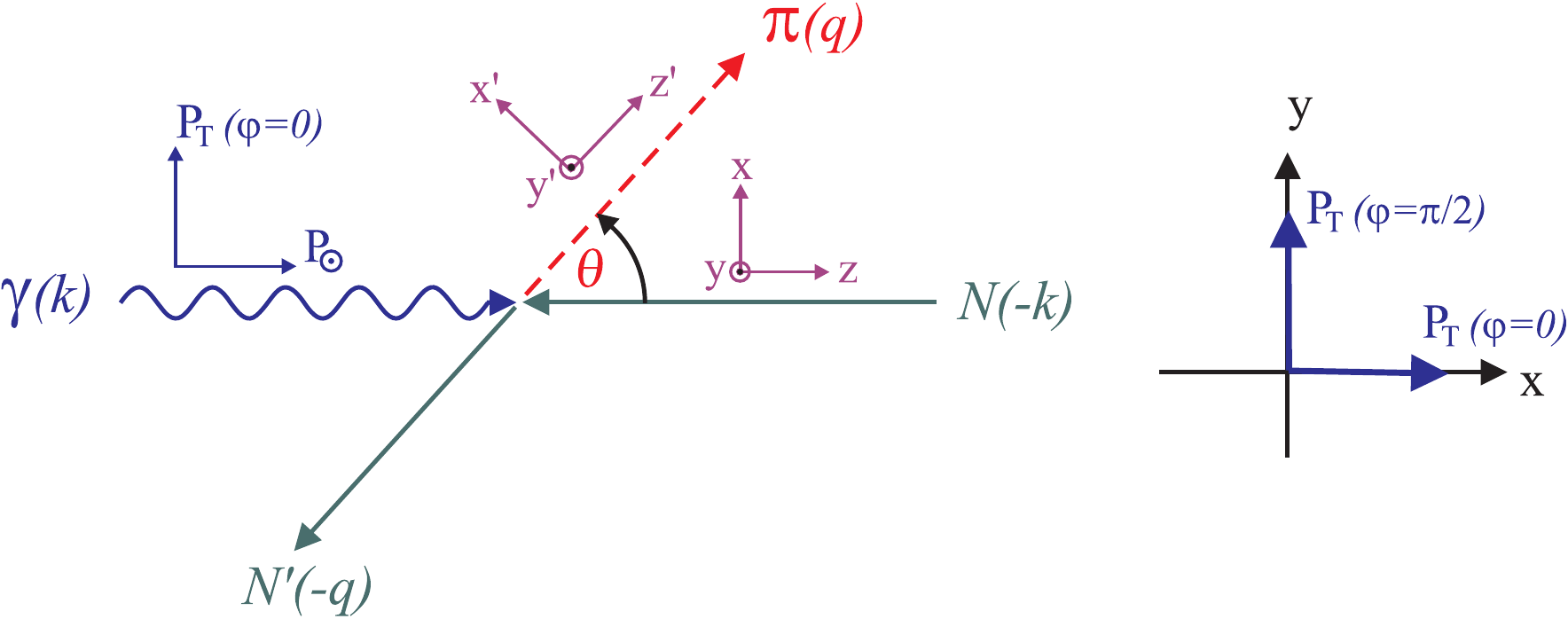} }
\caption{\label{fig:frames} Frames for polarization vectors in the
CM.}
\end{figure}

The polarized differential cross section can be classified into
three classes of double polarization experiments:\\
polarized photons and polarized target (types $(\mathcal S,\mathcal
{BT}$)
\begin{eqnarray}
\frac{d \sigma}{d \Omega} & = & \sigma_0\{ 1 - P_T \Sigma \cos 2
\varphi + P_x ( - P_T H \sin 2 \varphi + P_{\odot} F )
\nonumber \\
& & + P_y ( T - P_T P \cos 2 \varphi ) + P_z ( P_T G \sin 2 \varphi
- P_{\odot} E ) \} \; ,
\end{eqnarray}
polarized photons and recoil polarization (types $(\mathcal
S,\mathcal {BR}$)
\begin{eqnarray}
\frac{d \sigma}{d \Omega} & = & \sigma_0 \{ 1 - P_T \Sigma \cos 2
\varphi + P_{x'} ( -P_T O_{x'} \sin 2 \varphi - P_{\odot} C_{x'} )
\nonumber \\
& & + P_{y'} ( P - P_T T \cos 2 \varphi ) + P_{z'} ( -P_T O_{z'}
\sin 2 \varphi  - P_{\odot} C_{z'} ) \} \; ,
\end{eqnarray}
polarized target and recoil polarization (types $(\mathcal
S,\mathcal {TR}$)
\begin{equation}
\frac{d \sigma}{d \Omega}  =  \sigma_0 \{ 1 + P_{y} T + P_{y'} P +
P_{x'} ( P_x T_{x'} - P_{z} L_{x'} )  + P_{y'} P_y \Sigma + P_{z'}(
P_x T_{z'} + P_{z} L_{z'}) \} \; .
\end{equation}

In these equations $\sigma_0$ denotes the unpolarized differential
cross section, $\Sigma,T,P$ are single-spin asymmetries $(\mathcal
S)$, $E,F,G,H$ the beam-target asymmetries $(\mathcal B\mathcal T)$,
$O_{x'}$, $O_{z'}$, $C_{x'}$, $C_{z'}$ the beam-recoil asymmetries
$(\mathcal B\mathcal R)$ and $T_{x'},T_{z'},L_{x'},L_{z'}$ the
target-recoil asymmetries $(\mathcal T\mathcal R)$. The polarization
quantities are described in Fig.~1. The signs of the 16 polarization
observables of Eq.~(1,2,3) are in principle arbitrary, except for
the cross section $\sigma_0$, which is naturally positive. For the
15 asymmetries we use the sign convention of Barker et
al.~\cite{Barker75}, which is also used by the MAID and SAID partial
wave analysis groups. For other sign conventions, see
Ref.~\cite{Sandorfi:2011he}.

\subsection{Amplitude analysis}

Pseudoscalar meson photoproduction  has 8 spin degrees of freedom,
and due to parity conservation it can be described by 4 complex
amplitudes of 2 kinematical variables. Possible sets of amplitudes
are: Invariant amplitudes $A_i$, CGLN amplitudes $F_i$, helicity
amplitudes $H_i$ or transversity amplitudes $b_i$. All of them are
linearly related to each other and further combinations are
possible. Most often in the literature the helicity basis was chosen
and the 16 possible polarization observables can be expressed in
bilinear products
\begin{equation}\label{observables}
O_i(W,\theta) =
\frac{q}{k}\,\sum_{k,\ell=1}^4\,\alpha_{k,\ell}\,\,H_k(W,\theta)\,
H_l^*(W,\theta)\,,
\end{equation}
where $O_1$ is the unpolarized differential cross section $\sigma_0$ and all
other observables are products of asymmetries with $\sigma_0$, for details see
Table~\ref{tab:obs}.

From a complete set of 8 measurements $\{O_i(W,\theta)\}$ one can
determine the moduli of the 4 amplitudes and 3 relative phases. But
there is always an unknown overall phase, e.g. $\phi_1(W,\theta)$,
which can not be determined by additional measurements. This is,
however, not a principal problem as with the principally
undetermined phase of a quantum mechanical wave function. Already in
1963 Goldberger et al.~\cite{Goldberger:1963} discussed a method
using the idea of a Hanbury-Brown and Twiss experiment, and very
recently in 2012, Ivanov~\cite{Ivanov:2012na} discussed another
method using vortex beams to measure the phase of a scattering
amplitude. Both methods, however, are highly impractical for a meson
photoproduction experiment.

Therefore, the complete information is contained in a set of 4
reduced amplitudes,
\begin{equation}
\tilde{H_i}(W,\theta) =  {H_i}(W,\theta)\;e^{-i\,\phi_1(W,\theta)}
\end{equation}
of which $\tilde{H_1}$ is a real function, the others are complex,
resulting in a total of 7 real values for any given $W$ and
$\theta$.

\begin{table}[ht]
\caption{\label{tab:obs}Spin observables for pseudoscalar meson
photoproduction involving beam, target and recoil polarization in 4
groups, ${\mathcal S,\mathcal BT,\mathcal BR,\mathcal TR}$. A phase
space factor $q/k$ has been omitted in all expressions and the
asymmetries are given by $A=\hat{A}/\sigma_0$. In column 2 the
observables are expressed in terms of the Walker helicity
amplitudes~\cite{Walker} and in column 3 in $\sin\theta$ and
$x=\cos\theta$ with the leading terms for an $S,P$ wave truncation.
}
\vspace{0mm}
\begin{center}
\begin{tabular}{|c|c|c|}
\hline
 Spin Obs & Helicity Representation  & Partial Wave Expansion \\
\hline
$\sigma_0$    &$\frac{1}{2}(|H_1|^2+|H_2|^2+|H_3|^2+|H_4|^2)$   & $A_0^\sigma+A_1^\sigma x+A_2^\sigma x^2 + \cdots$\\
$\hat{\Sigma}$& Re$(H_1 H_4^* - H_2 H_3^*)$                           & $\sin^2\theta(A_0^\Sigma + \cdots)$\\
$\hat{T}$     & Im$(H_1 H_2^* + H_3 H_4^*)$                           & $\sin\theta(A_0^T+A_1^T x + \cdots)$\\
$\hat{P}$     & $-$Im$(H_1 H_3^* + H_2 H_4^*)$                        & $\sin\theta(A_0^P+A_1^P x + \cdots)$\\
\hline
$\hat{G}$     & $-$Im$(H_1 H_4^* + H_2 H_3^*)$                        & $\sin^2\theta(A_0^G+ \cdots)$ \\
$\hat{H}$     & $-$Im$(H_1 H_3^* - H_2 H_4^*)$                        & $\sin\theta(A_0^H+A_1^H x + \cdots)$\\
$\hat{E}$     &$\frac{1}{2}(-|H_1|^2+|H_2|^2-|H_3|^2+|H_4|^2)$ & $A_0^E+A_1^E x+A_2^E x^2 + \cdots$\\
$\hat{F}$     & Re$(H_1 H_2^* + H_3 H_4^*)$                           & $\sin\theta(A_0^F+A_1^F x + \cdots)$\\
\hline
$\hat{O_{x'}}$  & $-$Im$(H_1 H_2^* - H_3 H_4^*)$                      & $\sin\theta(A_0^{O_{x'}}+A_1^{O_{x'}}x+A_2^{O_{x'}}x^2+ \cdots)$ \\
$\hat{O_{z'}}$  & Im$(H_1 H_4^* - H_2 H_3^*)$                         & $\sin^2\theta(A_0^{O_{z'}}+A_1^{O_{z'}}x+\cdots)$\\
$\hat{C_{x'}}$  & $-$Re$(H_1 H_3^* + H_2 H_4^*)$                      & $\sin\theta(A_0^{C_{x'}}+A_1^{C_{x'}}x+A_2^{C_{x'}}x^2 + \cdots)$\\
$\hat{C_{z'}}$  &$\frac{1}{2}(-|H_1|^2-|H_2|^2+|H_3|^2+|H_4|^2)$      & $A_0^{C_{z'}}+A_1^{C_{z'}}x+A_2^{C_{z'}}x^2+A_3^{C_{z'}}x^3 + \cdots$\\
\hline
$\hat{T_{x'}}$  & Re$(H_1 H_4^* + H_2 H_3^*)$                      & $\sin^2\theta(A_0^{T_{x'}}+A_1^{T_{x'}}x+\cdots)$ \\
$\hat{T_{z'}}$  & Re$(H_1 H_2^* - H_3 H_4^*)$                         & $\sin\theta(A_0^{T_{z'}}+A_1^{T_{z'}}x+A_1^{T_{z'}}x^2+\cdots)$\\
$\hat{L_{x'}}$  & $-$Re$(H_1 H_3^* - H_2 H_4^*)$                      & $\sin\theta(A_0^{L_{x'}}+A_1^{L_{x'}}x+A_2^{L_{x'}}x^2+\cdots)$\\
$\hat{L_{z'}}$  &$\frac{1}{2}(|H_1|^2-|H_2|^2-|H_3|^2+|H_4|^2)$      & $A_0^{L_{z'}}+A_1^{L_{z'}}x+A_2^{L_{z'}}x^2+A_3^{L_{z'}}x^3+\cdots$\\
\hline
\end{tabular}
\end{center}
\end{table}

\begin{figure}[htb]
\centerline{
\includegraphics[width=0.36\textwidth]{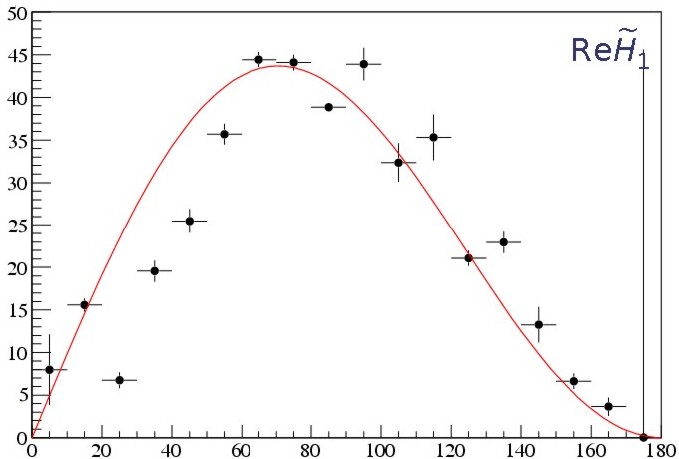}
\hspace*{0.8cm}
\includegraphics[width=0.36\textwidth]{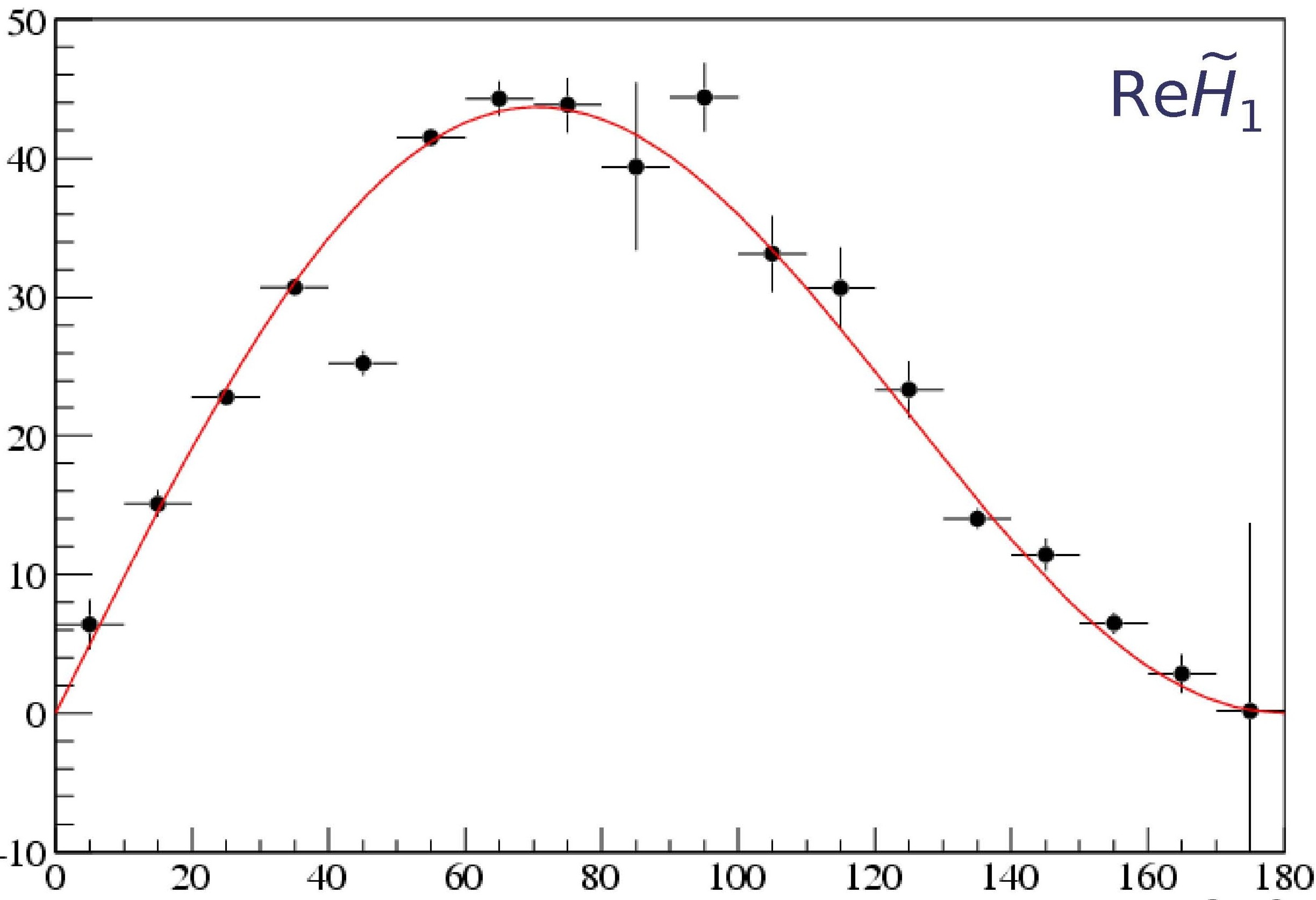}}
\vspace*{0.3cm} \centerline{
\includegraphics[width=0.36\textwidth]{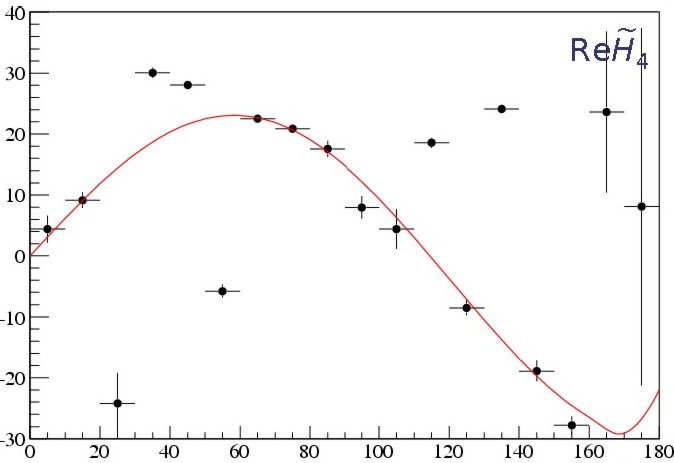}
\hspace*{0.8cm}
\includegraphics[width=0.36\textwidth]{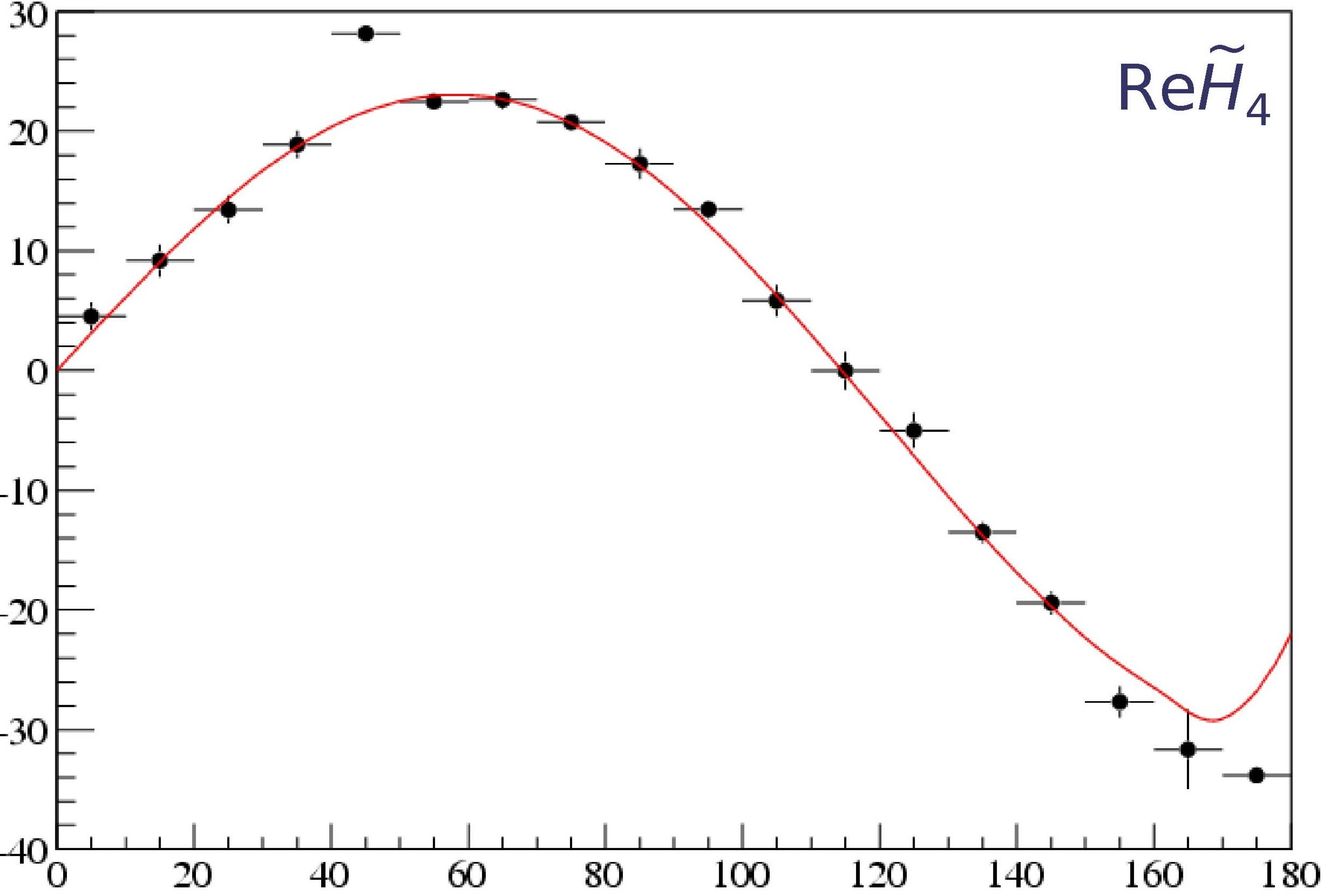}}
\caption{\label{fig:reduced_amplitudes} Comparison of the reduced
helicity amplitudes Re$\tilde{H}_1$ and Re$\tilde{H}_4$ between a
pseudo-data analysis with a complete dataset of 8 observables:
$\sigma_0,\Sigma,T,P,E,G,O_{x'},C_{x'}$ (left 2 panels) and with an
overcomplete dataset of 10 observables with additional $F,H$ (right
2 panels) for $\gamma p\rightarrow\pi^0 p$ at $E=320$ MeV as a
function of the c.m. angle $\theta$. The solid red curves show the
MAID2007 solutions. Amplitudes are in units of $10^{-3}/m_{\pi^+}$.}
\end{figure}

Figure 2 shows two of such amplitude analyses with a complete set of
8 observables and an overcomplete set of 10 observables. The data
used for this analysis has been generated as pseudo-data from
Monte-Carlo events according to the Maid2007 solution, see Sect. 3.
The figure shows the real parts of two out of four reduced helicity
amplitudes, Re$\tilde{H}_1$ and Re$\tilde{H}_4$. While the solution
with the complete set of 8 observables results in a rather bad
description of the true amplitudes, the solution of the overcomplete
set gives a satisfactory result.

\subsection{Truncated partial wave analysis}

Even with the help of unitarity in form of Watson's  theorem, the
angle-dependent phase $\phi_1(W,\theta)$ cannot be provided. This
has very strong consequences, namely a partial wave decomposition
would lead to wrong partial waves, which would be useless for
nucleon resonance analysis. It becomes obvious in the following
schematic formula
\begin{equation}
f_\ell(W)=\frac{2}{2\ell+1}\int\tilde{H}(W,\theta)e^{i\phi(W,\theta)}
P_\ell(\cos\theta)\;d\cos\theta\,, \label{phase_problem}
\end{equation}
where the desired partial wave $f_\ell(W)$ cannot be obtained from
the reduced helicity amplitudes $\tilde{H}(W,\theta)$ alone, as long
as the angle dependent phase $\phi(W,\theta)$ is unknown.

Our main goal in the data analysis of photoproduction is the search
for nucleon resonances and their properties. To better reach this
goal, one can directly perform a partial wave analysis from the
observables without going through the underlying helicity
amplitudes. Such an analysis would be a truncated partial wave
analysis (TPWA) with a minimal model dependence (i) from the
truncation of the series at a maximal angular momentum $\ell_{max}$
and (ii) from an overall unknown phase as in the case of the
amplitude analysis in the previous paragraph. However, in the TPWA
the overall phase would be only a function of energy and with
additional theoretical help it can be constrained without strong
model assumptions. Such a concept was already discussed and applied
for $\gamma,\pi$ in the 80s by Grushin~\cite{grushin} for a PWA in
the region of the $\Delta(1232)$ resonance.

Formally, the truncated partial wave analysis can be performed in
the following way. All observables can be expanded either in a
Legendre series or in a $\cos\theta$ series
\begin{eqnarray}
O_i(W,\theta) &=& \frac{q}{k}\,\,
sin^{\alpha_i}\theta\,\sum_{k=0}^{2\ell_{max}+\beta_i}
A_k^i(W)\,\,cos^k\theta\,, \label{eq:expans0}\\
A_k^i(W) &=& \sum_{\ell,\ell'=0}^{\ell_{max}}\; \sum_{k,k'=1}^4
\alpha_{\ell,\ell'}^{k,k'}\,\mathcal M_{\ell,k}(W)\, \mathcal
M_{\ell',k'}^*(W)\,, \label{eq:expans0b}
\end{eqnarray} where $k,k'$ denote the 4
possible electric and magnetic multipoles for each $\pi N$ angular
momentum $\ell\ge 2$, namely $\mathcal
M_{\ell,k}=\{E_{\ell+},E_{\ell-},M_{\ell+},M_{\ell-}\}$. For an
$S,P$ truncation $(\ell_{max}=1)$ there are 4 complex multipoles
$E_{0+},E_{1+},M_{1+},M_{1-}$ leading to 7 free real parameters and
an arbitrary phase, which can be put to zero for the beginning. In
Table~\ref{tab:obs} we list the expansion coefficients for all
observables that appear in an $S,P$ wave expansion. Already from the
8 observables of the first two groups $(\mathcal S,\mathcal BT)$ one
can measure a set of 16 coefficients, from which we only need 8 well
selected ones for a unique mathematical solution. This can be
achieved by a measurement of the angular distributions of only 5
observables, e.g. $\sigma_0,\Sigma,T,P,F$ or
$\sigma_0,\Sigma,T,F,G$. In the first example one gets even 10
coefficients, from which e.g. $A_1^P$ and $A_0^F$ can be omitted. In
the second case, there are 9 coefficients, of which $A_0^F$ can be
omitted. In practise one can select those coefficients, which have
the smallest statistical errors, and therefore, the biggest impact
for the analysis by keeping in mind that all discrete ambiguities
are resolved.

As has been shown by Omelaenko~\cite{omel} the same is true for any
PWA with truncation at $\ell_{max}$. For the determination of the
$8\ell_{max}-1$ free parameters one has the possibility to measure
$(8\ell_{max},\,8\ell_{max},\,8\ell_{max}+4,\,8\ell_{max}+4)$
coefficients for types $(\mathcal S,\mathcal {BT},\mathcal
{BR},\mathcal {TR})$, respectively.

\section{Partial wave analysis with pseudo-data}

In a first numerical attempt towards a model-independent partial
wave analysis, a procedure similar to the second method, the TPWA,
described above, has been applied~\cite{Workman:2011hi}, and
pseudo-data, generated for $\gamma,\pi^0$ and $\gamma,\pi^+$ have
been analyzed.

Events were generated over an energy range from $E_{lab}=200-1200$~MeV and a
full angular range of $\theta=0-180^\circ$ for beam energy bins of $\Delta
E_{\gamma}=10$~MeV and angular bins of $\Delta\theta=10^\circ$, based on the
MAID2007 model predictions~\cite{Drechsel:2007if}. For each observable,
typically $5\cdot10^6$ events have been generated over the full energy range.
\begin{figure}
\centerline{
\includegraphics[width=0.30\textwidth, keepaspectratio, angle=90]{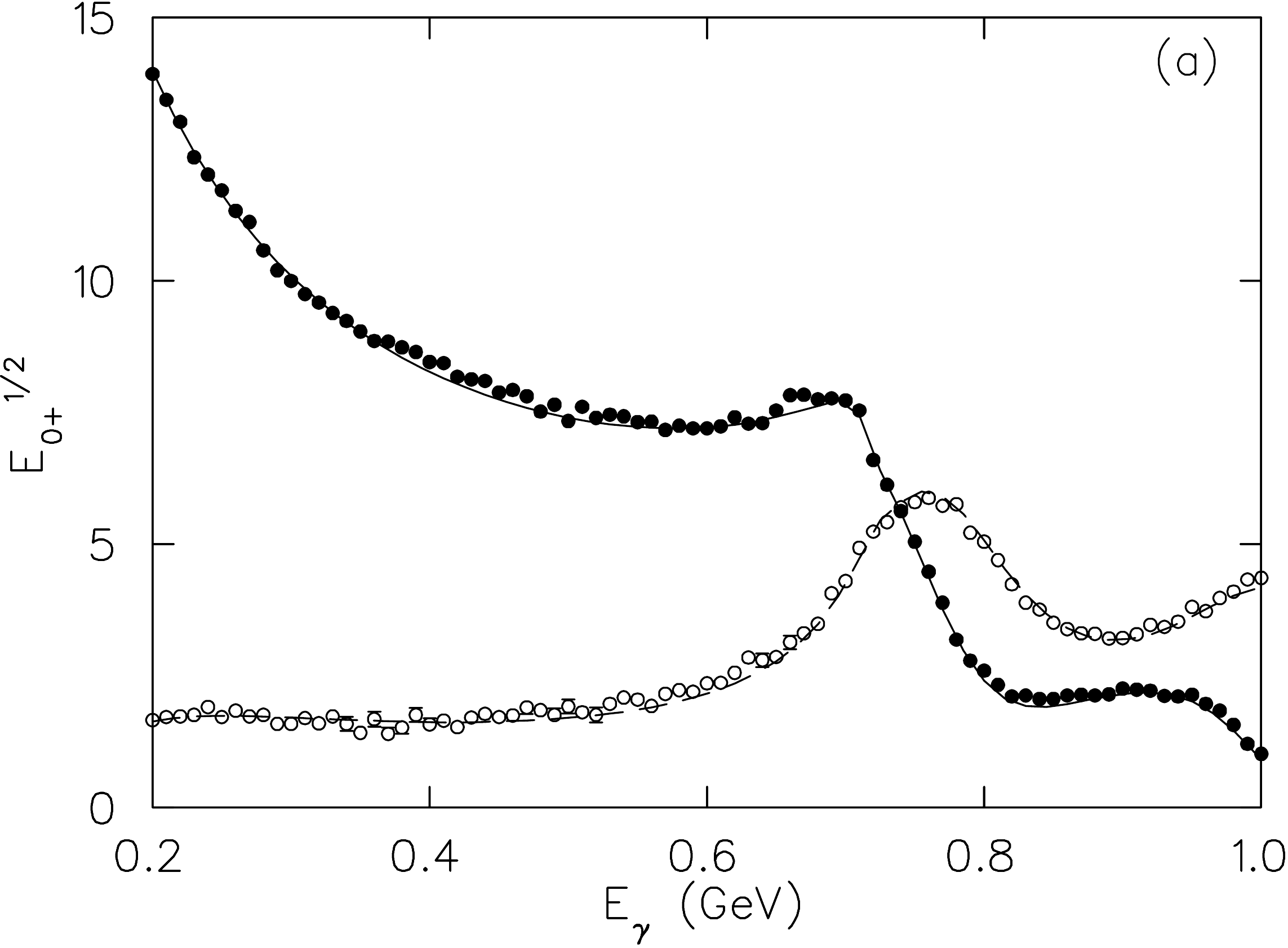}
\hspace*{0.8cm}
\includegraphics[width=0.30\textwidth, keepaspectratio, angle=90]{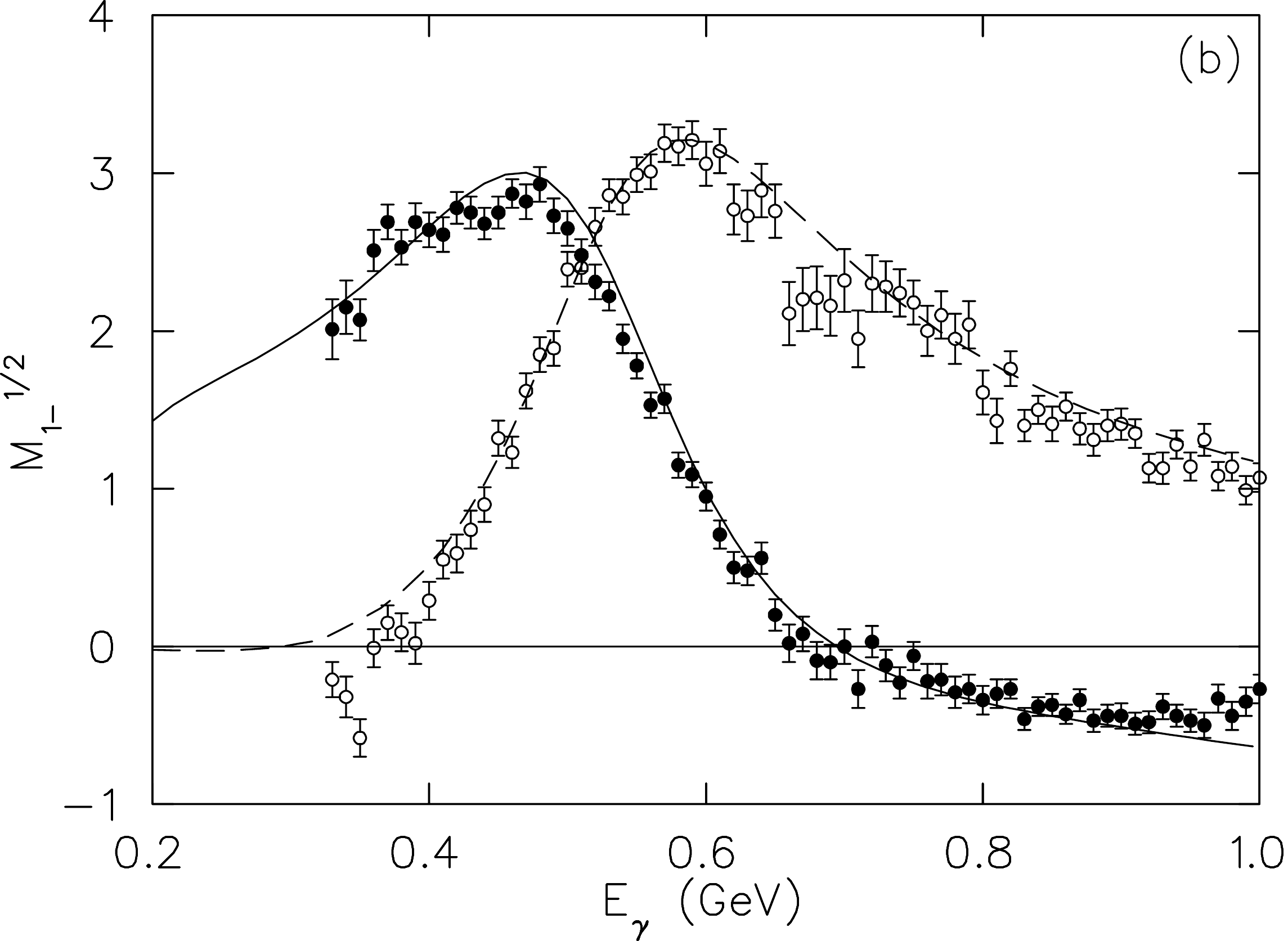}
} \caption{\label{fig:p11s11} Real and imaginary parts of (a) the
$S_{11}$ partial wave amplitude $E_{0+}^{1/2}$ and (b) the $P_{11}$
partial wave amplitude $M_{1-}^{1/2}$. The solid (dashed) line shows
the real (imaginary) part of the MAID2007 solution, used for the
pseudo-data generation.  Solid (open) circles display real
(imaginary) single-energy fits (SE6p) to the following 6 observables
without any recoil polarization measurement: $d\sigma/d\Omega$, two
single-spin observables $\Sigma$, $T$ and three beam-target double
polarization observables $E$, $F$, $G$. Multipoles are in millifermi
units.}
\end{figure}
For each energy bin a single-energy (SE) analysis has been performed
using the SAID PWA tools~\cite{Arndt:1989ww}.

A series of fits, SE4p, SE6p and SE8p have been
performed~\cite{Workman:2011hi} using 4, 6 and 8 observables,
respectively. Here the example using 6 observables
$(\sigma_0,\Sigma,T,E,F,G)$ is demonstrated, where no recoil
polarization has been used. As explained before, such an experiment
would be incomplete in the sense of an `amplitude analysis', but
complete for a truncated partial wave analysis. In
Fig.~\ref{fig:p11s11} two multipoles $E_{0+}^{1/2}$  and
$M_{1-}^{1/2}$ for the $S_{11}$ and $P_{11}$ channels are shown and
the SE6p fits are compared to the MAID2007 solution. The fitted SE
solutions are very close to the MAID solution with very small
uncertainties for the $S_{11}$ partial wave. For the $P_{11}$
partial wave we obtain a larger statistical spread of the SE
solutions. This is typical for the $M_{1-}^{1/2}$ multipole, which
is generally much more difficult to obtain with good
accuracy~\cite{Drechsel:2007if}, because of the weaker sensitivity
of the observables to this magnetic multipole. But also this
multipole can be considerably improved in an analysis with 8
observables~\cite{Workman:2011hi}.

\section{Summary and conclusions}

It is shown that for an analysis of $N^*$ resonances, the amplitude
analysis of a complete experiment is not very useful, because of an
unknown energy and angle dependent phase that can not be determined
by experiment and can not be provided by theory without a strong
model dependence. However, the same measurements or even less will
be very useful for a truncated partial wave analysis with minimal
model dependence due to truncations and extrapolations of Watson's
theorem in the inelastic energy region. A further big advantage of
such a PWA is a different counting of the necessary polarization
observables, resulting in very different sets of observables. While
it is certainly helpful to have polarization observables from 3 or 4
different types, for a mathematical solution of the bilinear
equations one can find minimal sets of only 5 observables from only
2 types, where either a polarized target or recoil polarization
measurements can be completely avoided.
\\


I would like to thank R. Workman, M. Ostrick and S. Schumann for
their contributions to this ongoing work. I want to thank the
Deutsche Forschungsgemeinschaft for the support by the Collaborative
Research Center 1044.


\begin{thebibliography}{9}

\bibitem{Barker75} I.~S. Barker, A. Donnachie, J.~K. Storrow, Nucl. Phys. B {\bf 95}, 347 (1975).

\bibitem{FTS92} C.~G. Fasano, F. Tabakin, B. Saghai, Phys. Rev. C {\bf 46}, 2430 (1992).

\bibitem{Keaton:1996pe}G.~Keaton and R.~Workman, Phys.\ Rev.\  C {\bf 54}, 1437 (1996).

\bibitem{chiang}W.-T. Chiang and F. Tabakin, Phys. Rev. C {\bf 55}, 2054 (1997).

\bibitem{Workman:2010xc}R.~L.~Workman, Phys.\ Rev.\  C {\bf 83}, 035201 (2011).

\bibitem{Workman:2011hi}
  R.~L.~Workman, M.~W.~Paris, W.~J.~Briscoe, L.~Tiator, S.~Schumann, M.~Ostrick and S.~S.~Kamalov,
  Eur.\ Phys.\ J.\ A {\bf 47}, 143 (2011).

\bibitem{Dey:2010fb}B.~Dey, M.~E.~McCracken, D.~G.~Ireland, C.~A.~Meyer, Phys.\ Rev.\  C {\bf 83}, 055208 (2011).

\bibitem{Sandorfi:2010uv}A.~M.~Sandorfi, S.~Hoblit, H.~Kamano, T.~-S.~H.~Lee,
  J.\ Phys.\ G {\bf 38}, 053001 (2011).

\bibitem{Sandorfi:2011he}
  A.~M.~Sandorfi, B.~Dey, A.~Sarantsev, L.~Tiator and R.~Workman,
  AIP Conf.\ Proc.\  {\bf 1432}, 219 (2012).

\bibitem{Goldberger:1963}
  M.~L.~Goldberger, H.~W.~Lewis and K.~M.~Watson,
  Phys.\ Rev.\  {\bf 132}, 2764 (1963).

\bibitem{Ivanov:2012na}
  I.~P.~Ivanov,
  Phys.\ Rev.\ D {\bf 85}, 076001 (2012).

\bibitem{Walker} R.~L. Walker, Phys. Rev. {\bf 182}, 1729 (1969).

\bibitem{grushin} V.~F. Grushin, in {\it Photoproduction of Pions on Nucleons and Nuclei}, edited
by A.~A. Komar, (Nova Science, New York, 1989), p. 1ff.

\bibitem{omel}A.~S. Omelaenko, Sov. J. Nucl. Phys. {\bf 34}, 406 (1981).

\bibitem{Drechsel:2007if}D.~Drechsel, S.~S.~Kamalov, L.~Tiator,  Eur.\ Phys.\ J.\  A {\bf 34}, 69 (2007).

\bibitem{Arndt:1989ww} R.~A.~Arndt, R.~L.~Workman, Z.~Li {\it et al.}, Phys.\ Rev.\  C {\bf 42}, 1853 (1990).

\end{thebibliography}
\end{document}